\documentclass[useAMS,usenatbib]{mn2e}

\usepackage{amstext}
\usepackage{amsmath, float}
\usepackage[utf8x]{inputenc}
\usepackage{graphicx}
\usepackage{bm}
\usepackage{booktabs}

\newcommand{\varv}{v}

\newcommand{\be}{\begin{equation}}
\newcommand{\ee}{\end{equation}}
\newcommand{\bdm}{\begin{displaymath}}
\newcommand{\edm}{\end{displaymath}}
\newcommand{\bea}{\begin{eqnarray}}
\newcommand{\eea}{\end{eqnarray}}
\newcommand{\ba}{\begin{align}}
\newcommand{\ea}{\end{align}}

\newcommand       \aap          {A\&A }
\newcommand       \ssr           {Space Sci. Rev.}
\newcommand       \aapr         {A\& A Rev.}
\newcommand       \nat         {Nature}

\newcommand       \apj        {ApJ}
\newcommand       \apjl        {ApJL}
\newcommand       \mnras   {MNRAS}

\newcommand       \jcap           {J. Cosm. Astrop. Phys.}
\newcommand       \physrep  {Phys. Rep.}
\newcommand       \prc         {Phys. Rev. C}
\newcommand       \prd         {Phys. Rev. D}

\title[Covariant and $3+1$ equations for DC-GRMHD]
{Covariant and $3+1$ Equations for Dynamo-Chiral \\ General Relativistic Magnetohydrodynamics}
\author[L. Del Zanna, N. Bucciantini] 
{L. Del Zanna$^{1,2,3}$\thanks{E-mail: luca.delzanna@unifi.it}, 
N. Bucciantini$^{2,1,3}$ \\
$^{1}$Dipartimento di Fisica e Astronomia, Universit\`a di Firenze, 
Via G. Sansone 1, 50019 Sesto Fiorentino (Firenze), Italy\\
$^{2}$INAF - Osservatorio Astrofisico di Arcetri, Largo E. Fermi 5, 50125 Firenze, Italy\\
$^{3}$INFN - Sezione di Firenze, Via G. Sansone 1, 50019 Sesto Fiorentino (Firenze), Italy}

\begin{document}

\date{Accepted ... Received ...}

\pagerange{\pageref{firstpage}--\pageref{lastpage}} \pubyear{2018}

\maketitle

\label{firstpage}

\begin{abstract}
The exponential amplification of initial seed magnetic fields in relativistic plasmas is a very important topic in astrophysics, from the conditions in the early Universe to the interior of neutron stars. While dynamo action in a turbulent plasma is often invoked, in the last years a novel mechanism of quantum origin has gained increasingly more attention, namely the \emph{Chiral Magnetic Effect} (CME). This has been recognized in semi-metals and it is most likely at work in the quark-gluon plasma formed in heavy-ion collision experiments, where the highest magnetic fields in nature, up to $B\sim 10^{18}~G$, are produced. This effect is expected to survive even at large hydrodynamical/MHD scales and it is based on the chiral anomaly due to an imbalance between left- and right-handed relativistic fermions in the constituent plasma. Such imbalance leads to an electric current parallel to an external magnetic field, which is precisely the same mechanism of an $\alpha$-dynamo action in classical MHD. Here we extend the close parallelism between the chiral and the dynamo effects to relativistic plasmas and we propose a unified, fully covariant formulation of the generalized Ohm's law. Moreover, we derive for the first time the $3+1$ general relativistic MHD equations for a chiral plasma both in flat and curved spacetimes,  in view of  numerical investigation of the CME in compact objects, especially magnetars, or of the interplay among the non-ideal magnetic effects of dynamo, the CME and reconnection.
\end{abstract}


\section{Introduction}

The baryonic component of the Universe is found almost entirely in the form of plasma, typically ionized gas for the standard conditions of the heliosphere or of the interstellar medium. On the largest (hydrodynamical) scales astrophysical plasmas can be safely treated as an electric conductive fluid, locally neutral, where currents and magnetic fields play an important role (\emph{magnetohydrodynamics}, or MHD). This approximation can be either employed in its classical version (non-relativistic speed and temperature), or in the most general case appropriate to a relativistic fluid. In the latter case the theory is named relativistic MHD, or GRMHD when the presence of strong gravity requires the use of general relativity (e.g. for compact objects like neutron stars and black holes). 

The magnetic fields involved in astrophysical plasmas may be extremely strong and they are responsible for many steady and variable emission processes, such as the coronal activity and the solar flares driven by sub-photospheric motions and flux emergence \citep{Priest:2002,Shibata:2011}, the electromagnetic spin-down emission from pulsars \citep{Pacini:1968,Spitkovsky:2006,Philippov:2015}, the magnetar flares \citep{Lyutikov:2006a}, and the launch of relativistic outflows and jets either in active galactic nuclei (AGNs) \citep{Blandford:1977,Komissarov:2004,Hawley:2006,Mignone:2010b,Tchekhovskoy:2016a} or in the progenitors of gamma-ray burst (GRBs) \citep{Aloy:2000,Drenkhahn:2002,Barkov:2008,Bucciantini:2009,Komissarov:2009,Rezzolla:2011}.

Given that ideal MHD alone fails at explaining the process of magnetization of a fluid initially with $\bm{B}=0$, one of the most important problems for astrophysical plasmas is the origin of the constituent magnetic fields, in the various environments outlined above, and their growth through cosmic time. In stars such fields are likely to be originated via the so-called $\alpha-\Omega$ \emph{dynamo processes} \citep{Moffatt:1978}, a mechanism capable to amplify initial tiny seed fields exponentially and giving rise to the observed magnetic variable activity (e.g. the solar cycle). Similar mechanisms generated and maintained by turbulence are also invoked to explain the magnetic fields in accretion disks and in the interstellar medium \citep{Brandenburg:2005}. As far as the generation of the primordial magnetic field is concerned, this must have been originated during or right after the Big Bang, due to inflation or phase transitions \citep{Turner:1988,Sigl:1997,Giovannini:2004,Kandus:2011,Campanelli:2013,Kahniashvili:2013}. These fields should be able to survive cosmological expansion provided that the observation of lower limits in the intergalactic medium (IGM), $B > 3\times 10^{-16}$~G \citep{Neronov:2010}, are indeed an indication of a primordial origin. Alternative explanations for the IGM magnetic fields are aperiodic plasma fluctuations \citep{Schlickeiser:2012} or battery processes based on photoionization during the reionization epoch \citep{Durrive:2015,Durrive:2017}. 

A promising mechanism, first proposed precisely for the origin of primordial magnetic fields, is based on the creation of an equilibrium current arising in a \emph{chiral} system of charged fermions where parity conservation is violated \citep{Vilenkin:1980, Vilenkin:1982}. This is an intrinsically quantum phenomenon and it has been later rediscovered using different arguments \citep[e.g.][]{Alekseev:1998}. In the context of the \emph{Quark-Gluon Plasma} (QGP) produced by heavy-ion collisions this phenomenon is known as the \emph{Chiral Magnetic Effect} \citep{Kharzeev:2008,Fukushima:2008,Kharzeev:2014}. This quantum effect arises when there is a chiral imbalance between right- and left-handed fermions leading to the creation of an electric current along an external magnetic field, that is $\bm{J}_\mathrm{CME}\propto\bm{B}$. This field-aligned current is topologically protected even in the presence of strong interactions, thus it is of \emph{non-dissipative} nature and expected to survive at a macroscopic level, and hence to affect hydrodynamic properties like the transport coefficients \citep{Son:2009,Kharzeev:2011,Huang:2016}. 

Experimental evidence of the CME has been recently recognized in condensed matter physics in the so-called \emph{Weyl semi-metals}  \citep{Xiong:2015,Huang:2015,Li:2016}. As far as QGP is concerned, the interplay of quantum anomalies with both magnetic field and vorticity leads to a variety of phenomena, most notably the CME itself, which are under investigation at the Relativistic Heavy Ion Collider (RHIC) and the Large Hadron Collider (LHC). Promising progress has been achieved but more theoretical and experimental work is needed to unambiguously identify the CME and related phenomena \citep{Kharzeev:2016}. A macroscopic quantum effect which has been clearly identified is the coupling of particles spin with the QGP fluid vorticity (rather than the magnetic field), leading to hyperons global polarization. This was first observed in relativistic viscous hydrodynamical simulations \citep{Becattini:2015}, and recently confirmed in non-central Au-Au collisions at RHIC \citep{STAR-coll:2017}. 

In addition, ultrarelativistic collisions of heavy ions are known to produce the highest magnetic fields ever measured, up to $B\sim 10^{18}~G$, even though these are expected to decay very fast during the fireball expansion even in the presence of a plasma \citep{Tuchin:2013,McLerran:2014,Pang:2016,Pu:2016}, therefore, it is natural to identify the QGP as the best environment where to look for CME evidences \citep{Tuchin:2015,Li:2016a}. More realistic calculations based on numerical simulations of the expansion of an initially strongly magnetized QGP fireball in the relativistic MHD regime are starting to appear in the literature \citep{Inghirami:2016,Roy:2017,Das:2017,Inghirami:2018}.

As far as astrophysical applications are concerned, the importance of the CME is that the resulting field-aligned current is expected to act like a dynamo and amplify seed magnetic fields in several environments. The origin of primordial magnetic fields may be attributed to the CME provided a large reservoir of chiral asymmetry was present and temperature remains high enough, above $T \simeq10$~MeV \citep{Boyarsky:2012,Tashiro:2012}, whereas the formation of a turbulent spectrum has been recently investigated numerically \citep{Dvornikov:2017,Brandenburg:2017,Schober:2018}. It is found that after the initial CME instability, leading to magnetic field growth and driven turbulence, a new mean-field dynamo effect arises and the field keeps on increasing of several orders of magnitude, before the final saturation stage to the observed values of magnetization. Similar effects may act in the interior of a \emph{proto-neutron star} \citep{Pons:1999}, where due to the highly turbulent and hot ($T>100$~MeV) medium the magnetic field may be easily amplified to values beyond the quantum threshold ($B_\mathrm{QED} \approx 4.4\times 10^{13}$~G), either by dynamo processes \citep{Duncan:1992}, or by the CME itself due to the chiral imbalance produced in the URCA processes during deleptonization \citep{Dvornikov:2015,Sigl:2016,Yamamoto:2016}. Once this extremely strong field has emerged in the star corona, depending on the field topology the typical manifestations of \emph{magnetar} activity may arise \citep{Turolla:2015}.

In spite of the wealth of potential applications of the CME to relativistic plasmas, the theoretical picture where this quantum effect is self-consistently included within the MHD framework, as appropriate to the macroscopic fluid scales, is not well established yet. The CME current is either included \emph{by hand} in the Maxwell equations, retrieving the so-called Chern-Simons-Maxwell set of equations \citep{Tuchin:2015,Qiu:2017}, or the approach of statistical mechanics based on the second law of thermodynamics is followed \citep{Kharzeev:2011,Boyarsky:2015,Yamamoto:2016a,Giovannini:2016}. In any case most of the theoretical frameworks proposed so far, either treat the equations in a covariant form but then remain in the reference frame of the fluid, or evolve the full system for the anomaly coefficient, the hydrodynamical quantities, and the magnetic field (including the interplay with turbulence) within non-relativistic MHD \citep{Pavlovic:2017,Rogachevskii:2017}. 

What is still missing is a chiral MHD theory valid for fully relativistic plasmas, including the possible presence of relativistic bulk motions and/or strong gravity, to be treated by solving the Einstein field equations, for instance for applications to accretion onto black holes or for neutron star interior. A similar approach was presented in the case of the mean-field dynamo action by \cite{Bucciantini:2013}, where the upgraded version of the \texttt{ECHO} (\emph{Eulerian Conservative High-Order}) \citep{Del-Zanna:2007,Bucciantini:2011a} numerical code for non-ideal GRMHD including the dynamo effect was described and tested in different geometries and spacetime metrics.

In the present paper we propose to fill this gap and we extend the strong analogy of the CME effect and the dynamo to the fully relativistic case. We first write the covariant form of Ohm's law for a dynamo-chiral plasma, including dissipation, in the comoving frame of the fluid. Then we transform this in the reference frame of the so-called \emph{Eulerian observer}, together with the evolution equations for electromagnetic fields and matter, allowing for the $3+1$ splitting of the equations as needed for numerical integration. We thus derive for the first time the equations for \emph{Dynamo-Chiral General Relativistic MagnetoHydroDynamics} (DC-GRMHD), valid either in Minkowski flat space or in any curved manifold of general relativity. Finally, we present a numerical application to the growth of magnetic fields due to the CME for a simplified magnetar model, assuming a conformally flat evolving spacetime metric.

\section{The Chiral Magnetic Effect}

As discussed in the Introduction (see the references therein for the relevant literature), the Chiral Magnetic Effect (CME) in a system composed by high-temperature chiral fermions is the phenomenon of electric charge separation, and hence creation of a current, along an external magnetic field. In spite of its quantum nature, being related to the imbalance of right- and left-handed chirality of particles with spin $1/2$ (for example in a pair plasma), the current is non-dissipative and the effect is macroscopic and it is expected to affect the overall dynamics of the plasma itself. If $\bm{B}$ is an external magnetic field present in a plasma with chiral anomaly, the electric current induced by the resulting charge separation is 
\be
\bm{J}_\mathrm{CME} = \sigma_\mathrm{A} \bm{B}, 
\label{eq:cme}
\ee
where $\sigma_\mathrm{A}$ is the \emph{axial}, or \emph{chiral}, conductivity coefficient. This expression looks analogous to the usual one responsible for Ohmic dissipation, the standard Ohm's law being $\bm{J}_\mathrm{Ohm} = \sigma \bm{E}$, with the crucial difference that here the current is proportional to $\bm{B}$ rather than $\bm{E}$. 

In case both CME and Ohmic dissipation are present in the chiral plasma, the Maxwell equations (from now on we assume natural units with $4\pi\to 1,\,c\to 1,\,\hbar\to 1$)
\begin{align}
& \partial_t \bm{B}  + \bm{\nabla}\times\bm{E} = 0, \qquad \bm{\nabla}\cdot\bm{B} = 0 \nonumber \\
& \partial_t \bm{E}  - \bm{\nabla}\times\bm{B} = - \bm{J} , \qquad \bm{\nabla}\cdot\bm{E} = \rho_\mathrm{e}
\end{align}
where $ \rho_\mathrm{e}$ is the charge density (a derived quantity in MHD, or even set to zero assuming local neutrality at the hydrodynamical scales), are closed by assuming the generalized form of Ohm's law
\be
\bm{J} = \bm{J}_\mathrm{Ohm}  + \bm{J}_\mathrm{CME}  = \sigma\bm{E} + \sigma_\mathrm{A} \bm{B}.
\label{eq:ohmcme}
\ee
Note that the relation expressed in the above form is valid for non-relativistic velocities or in the comoving frame of the fluid. For constant conductivity coefficients the evolution equation for the magnetic field inside a chiral plasma is
\be
\partial_t^2\bf{B} + \sigma\,\partial_t\bf{B} =   \sigma_\mathrm{A} (\bm{\nabla}\times \bm{B}) +\nabla^2\bf{B} , 
\label{eq:Bevol}
\ee
and neglecting timescales shorter than $1/\sigma$, it is easy to see that the magnetic field may grow in time due to the CME, even exponentially, against Ohmic dissipation.

A peculiar aspect of CME is that Eq.~(\ref{eq:cme}) introduces a parity-violation otherwise absent in Maxwell's equations. Hence, this may happen only in a chiral medium with broken parity symmetry, since $\bm{B}$ is a parity-even pseudo-vector while the electric current is a parity-odd vector. Moreover, since both $\bm{J}$ and $\bm{B}$ are odd under time reversal, the chiral conductivity $\sigma_\mathrm{A}$ must be time-even, which is an unusual behavior, given that for Ohmic conductivity $\sigma$ is the opposite, since $\bm{E}$ is time-even. This is however typical of non-dissipative currents, for example those present in super-conducting media, so no entropy production is expected by the CME current \citep{Kharzeev:2011}.

Let us now investigate how the chiral conductivity depends on other quantities and how this can be evolved in time and coupled to the other MHD equations. If $n_\mathrm{A}$ and $\bm{J}_\mathrm{A}$ are the axial charge and current densities of the unbalanced fermions, their conservation is known to be broken by the presence of an electric field aligned to the magnetic field, so that the continuity equation becomes inhomogeneous. The resulting (\emph{anomalous}) evolution equation for the axial charge is usually written as \citep[e.g.][]{Kharzeev:2014,Rogachevskii:2017}
\be
\partial_t n_\mathrm{A} + \bm{\nabla}\cdot\bm{J}_\mathrm{A} = C_\mathrm{A} \bm{E}\cdot\bm{B},
\label{eq:axial}
\ee
where $C_\mathrm{A}=e^2/2\pi^2$ for charged massless fermions, such as a relativistically hot pair-plasma. By integrating in space and introducing the total time derivative, the axial charge produced in unit time in a volume $V$ where $\bm{E}\cdot\bm{B}\neq 0$ is
\be
\frac{dQ_\mathrm{A}}{dt} = C_\mathrm{A} \int_V \bm{E}\cdot\bm{B}  \, \mathrm{d}x^3, \qquad 
Q_\mathrm{A} = \int_V n_\mathrm{A} \, \mathrm{d}x^3.
\ee
Introduce now the \emph{chiral chemical potential} $\mu_\mathrm{A}$, conjugated to the axial charge $n_\mathrm{A}$, which is in practice the energy needed to produce a single anomaly.  The chiral conductivity in terms of $\mu_\mathrm{A}$ is simply
\be
\sigma_\mathrm{A} = C_\mathrm{A} \mu_\mathrm{A},
\label{eq:sigma}
\ee
where the same constant $C_\mathrm{A}$ appearing in Eq.~(\ref{eq:axial}) appears.

In the case when spatial variations of $n_\mathrm{A}$ are not strong and for a linearized equation of state $n_\mathrm{A} =\chi_\mathrm{A} \mu_\mathrm{A}$, where $\chi_\mathrm{A}$ is the chiral, or axial, \emph{susceptibility} (proportional to the square of the temperature, $\chi_\mathrm{A}\propto T^2$), one may write
\be
\mu_\mathrm{A} = \frac{n_\mathrm{A}}{\chi_\mathrm{A}} \simeq \frac{Q_\mathrm{A}}{\chi_\mathrm{A} V},
\ee
hence the chiral conductivity, and thus the CME current itself, are actually quadratic in the constant $C_\mathrm{A}$. In this simple case, the time evolution for $\mu_\mathrm{A}$ may be written as
\be
\frac{d\mu_\mathrm{A}}{dt} = \frac{1}{\chi_\mathrm{A} V} \frac{dQ_\mathrm{A}}{dt} =
\frac{C_\mathrm{A}}{\chi_\mathrm{A} V} \int_V  \bm{E}\cdot\bm{B}\, \mathrm{d}x^3 ,
\label{eq:muevol}
\ee
thus the growth of the CME current is clearly enhanced in the regions where $\bm{E}\cdot\bm{B}\neq 0$ of a non-ideal plasma. 

On the other hand, when spatial dependences of the chiral magnetic potential are not negligible, an evolution equation may be derived from Eq.~(\ref{eq:axial}) assuming that the axial current can be expressed as
\be
\bm{J}_\mathrm{A} = - D_\mathrm{A} \bm{\nabla} n_\mathrm{A}, 
\ee
where $D_\mathrm{A}$ is a (constant) diffusion coefficient. Employing $\chi_\mathrm{A}$ once again, the following convection-diffusion equation with a source proportional to the anomaly term is found \citep{Boyarsky:2015}
\be
\partial_t \mu_\mathrm{A} - D_\mathrm{A} \nabla^2 \mu_\mathrm{A}  = 
\frac{C_\mathrm{A}}{\chi_\mathrm{A}} \bm{E}\cdot\bm{B},
\ee
that may be solved to provide the time and space dependency of $\sigma_\mathrm{A}$, and hence $\bm{J}_\mathrm{CME}$, using Eqs.~(\ref{eq:sigma}) and (\ref{eq:cme}).

A very interesting property of the CME is the relation of the axial charge with the \emph{magnetic helicity} $\mathcal{H}$ of the plasma, for which we have \citep[e.g.][]{Biskamp:1993}
\be
\mathcal{H} =  \int \bm{A}\cdot\bm{B}  \, \mathrm{d}x^3, \qquad
\frac{d\mathcal{H} }{dt} = - 2\int_V  \bm{E}\cdot\bm{B}\, \mathrm{d}x^3 ,
\label{eq:helicity}
\ee
where $\bm{A}$ is the usual magnetic vector potential. Note that $\mathcal{H}$ looks gauge dependent, but if $\bm{B}$ is parallel to the external boundary, or the latter is far away in a region of vanishing magnetization, then $\mathcal{H}$ becomes gauge invariant. Now, the magnetic helicity is known to be preserved in an ideal plasma, even in the presence of motions at all scales, while it is allowed to vary on the (slow) diffusion timescales in three-dimensional reconnection events, where the magnetic field topology changes \citep{Berger:1984,Priest:2000,Blackman:2015}. Similarly here, in a chiral plasma $\mathcal{H}$ is not conserved, but we have the balance
\be
\frac{d}{dt}\left( Q_\mathrm{A} + \tfrac{1}{2} C_\mathrm{A}\, \mathcal{H} \right) = 0.
\ee
As a consequence, one may also rewrite Eq.~(\ref{eq:muevol}) as  \citep{Boyarsky:2012}
\be
\frac{d\mu_\mathrm{A}}{dt} = - \frac{C_\mathrm{A}}{2\chi_\mathrm{A} V} \frac{d \mathcal{H}}{dt}  ,
\label{eq:muevol2}
\ee
and the CME growth is now related to a decrease of magnetic helicity. The complex interplay between the chiral anomaly and magnetic helicity at different spatial scales leads to very important consequences on the chiral MHD turbulent (inverse) cascade  and magnetic field amplification \citep{Tashiro:2012,Hirono:2015,Pavlovic:2017,Rogachevskii:2017}, as recently confirmed also on the basis of numerical simulations \citep{Schober:2018}.

\section{The turbulent mean-field dynamo closure for MHD}

On the large scales of astrophysical sources, quantum anomalies and the CME in particular are usually neglected, though we have seen in the Introduction that these may be important for the early phases of cosmic expansion or inside the cores of (proto) neutron stars. The problem of magnetic field amplification has traditionally been addressed by invoking turbulence and the so-called \emph{mean-field dynamo} effect. 

When MHD quantities are decomposed into large-scale mean values and stochastic fluctuations, as appropriate in a turbulent medium, the Ohm law written in the comoving frame of the plasma for resistive (classical) MHD becomes
\be
\bm{E}^\prime \equiv \bm{E} + \bm{\varv}\times \bm{B} = - < \delta\bm{\varv}\times\delta\bm{B} > + \eta \bm{J}; \quad \bm{J} = \bm{\nabla} \times \bm{B},
\ee
where a (non-relativistic) bulk flow velocity $\bm{\varv}$ is allowed. Here $\eta\equiv 1/\sigma$ is the standard Ohmic resistivity coefficient (assumed to be a scalar, neglecting anisotropies for simplicity) and we retain the usual vector notation for the averaged fields. Now, the key assumption in turbulent dynamo theory is that the mean of the quadratic term, which is essentially an electromotive force, can be written as
\be
< \delta\bm{\varv}\times\delta\bm{B} > = \alpha_\mathrm{dyn} \bm{B} - \beta_\mathrm{dyn} (\bm{\nabla}\times\bm{B}),
\ee
so that a mean-field closure is reached, where the $\alpha_\mathrm{dyn}$ and $\beta_\mathrm{dyn}$ terms depend on the turbulent properties (namely turbulent fluid helicity, energy, and correlation time) \citep{Moffatt:1978,Parker:1979,Krause:1980}. Thus, in a turbulent plasma, the new form for Ohm's law is
\be
\bm{E}^\prime = - \alpha_\mathrm{dyn}\bm{B} + \eta\bm{J},
\label{eq:ohm1}
\ee
where the contribution of the turbulent diffusivity  $\beta_\mathrm{dyn}$ has been absorbed in the $\eta$ coefficient for simplicity. 

When combined to the Maxwell equations into the induction equation for $\bm{B}$, the $\alpha$-term leads to exponentially growing modes of the magnetic field (with a growth rate $\gamma\propto\alpha_\mathrm{dyn}$), the proper dynamo effect, whereas the second term leads to diffusion. Indeed, for constant $\alpha_\mathrm{dyn}$ and $\eta$ coefficients, one finds
\be
\partial_t \bm{B} = \bm{\nabla}\times (\bm{v}\times\bm{B}) + 
\alpha_\mathrm{dyn} (\bm{\nabla}\times\bm{B}) + \eta \bm{\nabla}^2\bm{B},
\label{eq:Bevol2}
\ee
and for a static plasma with $\bm{B}\sim\exp (ikx+\gamma t)$, the growth rate is exactly $\gamma=\alpha_\mathrm{dyn} k$ if diffusion is neglected. The exponential growth of the magnetic field is the main goal of the dynamo effect, but obviously in any realistic setup the back reaction on the MHD structure or magnetic diffusion will prevent the unlimited growth predicted by solving the induction equation alone. If one prefers to remain in the kinematical approach, the trick is usually to introduce a quenching effect of the kind
\be
\alpha_\mathrm{dyn} (\bm{B}) = \frac{\alpha_0}{1 + (\bm{B}/B_\mathrm{eq})^2},
\ee 
where $B_\mathrm{eq}$ is an equipartition field. In addition to the dynamo instabilities, the $\alpha$ term is also responsible for the propagation of dynamo waves, first predicted for the solar convection zone \citep{Parker:1955}.

An alternative form for the Ohm law in Eq.~(\ref{eq:ohm1}) can be written in terms of conduction coefficients, and here we chose the symmetric expression
\be
\bm{J} = \sigma_E \bm{E}^\prime + \sigma_B \bm{B},
\label{eq:ohm2}
\ee
where $\sigma_E=1/\eta$ and $\sigma_B = \alpha_\mathrm{dyn}/\eta$. The presence of a conduction current proportional to $\bm{B}$ itself is the distinctive characteristic of the $\alpha$-dynamo action leading to the exponentially growing modes discussed above. However, as we have shown in the previous section, just compare the above expression with Eq.~(\ref{eq:ohmcme}) or Eq.~(\ref{eq:Bevol}) when $\bm{\varv}=0$ with Eq.~(\ref{eq:Bevol2}), it is also typical of the CME, which in fact is known to yield the same type of growing modes, propagating waves, and turbulent cascade \citep[e.g.][]{Rogachevskii:2017}. Notice that contrary to the mean-field dynamo described here, the chiral mechanism operates even for simple laminar flows or non-helical turbulence. 

Therefore, from now on we propose a unified treatment of both the \emph{dynamo-chiral} effects within the MHD regime, in which $\sigma_B$ can be due either to the mean-field dynamo of classical MHD as defined above, or to the CME due to the presence of the chiral anomaly,  hence $\sigma_B=\sigma_\mathrm{A}$ as in Eq.~(\ref{eq:cme}) (or to both contributions, and in this case $\sigma_B$ is simply obtained by summing the two coefficients). On the other hand, $\sigma_E$ is always given by the inverse of the resistivity, either of Ohmic (collisional) type or due to the turbulent mean-field closure. When all forms of dissipation vanish ($\sigma_E\to \infty$) and the dynamo-chiral action can be neglected ($\sigma_B\to 0$), we retrieve the ideal MHD condition for infinite conductivity
\be
\bm{E}^\prime=\bm{E} + \bm{\varv}\times \bm{B} =0.
\ee

\section{Dynamo-chiral GRMHD equations}

So far we have summarized the main properties of the CME and of mean-field dynamo action as appropriate for non-relativistic plasmas, and we have proposed Eq.~(\ref{eq:ohm2}) as a universal Ohm's law valid for both effects, to be incorporated within the MHD system. As discussed in the Introduction, since the current research on the CME is mainly focused to the physics of the QGP plasma formed during heavy-ion collisions, only special relativistic treatments can be found in the literature. Having in mind applications relevant for Astrophysics, where the use of the Einstein theory of gravitation may be important, in the present section we derive for the first time the equations for the full system of \emph{Dynamo-Chiral General Relativistic MagnetoHydroDynamics} (DC-GRMHD) equations, valid for any curved spacetime metric, first in covariant form, and later moving to the $3+1$ formalism, as needed for applications of numerical relativity.

\subsection{Covariant formulation}

The equations for one-fluid GRMHD are the conservation laws for mass and total (matter and electromagnetic fields) energy-momentum 
\be
\nabla_\mu (\rho u^\mu)  =  0, \qquad \nabla_\mu T^{\mu\nu}  =  0,
\label{eq:hydro}
\ee
where $\rho$ is the mass density and $u^\mu$ is the fluid velocity (here we have assumed the so-called Eckart reference frame), and Maxwell's equations
\be
\nabla_\mu F^{\mu\nu} = - I^\nu, \quad \nabla_\mu F^{\star\mu\nu} = 0.
\label{eq:maxwell}
\ee
Here $F^{\mu\nu}$ is the Faraday tensor, $F^{\star\mu\nu}\equiv \textstyle{\frac{1}{2}}\epsilon^{\mu\nu\lambda\kappa}F_{\lambda\kappa}$ its dual, $I^\mu$ the four-current, satisfying the condition $\nabla_\mu I^\mu = 0$ for electric charge conservation, $\nabla_\mu$ is the covariant derivative associated to the four-metric $g_{\mu\nu}$, and $\epsilon^{\mu\nu\lambda\kappa}$ the Levi-Civita pseudo-tensor. 
Notice that here we have neglected possible polarization and magnetization effects of the plasma, therefore we do not make distinction between microscopic and macroscopic fields. While the total energy-momentum is conserved, we know that the electromagnetic fields act on the plasma via the Lorentz force, thus
\be
\nabla_\mu T^{\mu\nu} _\mathrm{m}=-\nabla_\mu T^{\mu\nu} _\mathrm{f}= - I_\mu\,F^{\mu\nu},
\label{eq:mf}
\ee
where $T^{\mu\nu} _\mathrm{m}$ is the \emph{matter} contribution and
\be
T^{\mu\nu} _\mathrm{f} = F^{\mu\lambda}F^\nu_{\,\lambda} - 
\tfrac{1}{4}g^{\mu\nu}F^{\lambda\kappa}F_{\lambda\kappa}
\ee
is the \emph{field} contribution to the energy-momentum tensor.

Let us now decompose all MHD quantities according to $u^\mu$. When the dissipative terms due to viscosity and heat conduction are negligible, the matter contribution to the energy-momentum tensor is simply provided by ideal hydrodynamics as 
\be
T_\mathrm{m}^{\mu\nu} = (\varepsilon + p)u^\mu u^\nu + p g^{\mu\nu},
\ee
where $\varepsilon = T_\mathrm{m}^{\mu\nu} u_\mu u_\nu$ is the fluid energy density, and $p$ is the pressure ($p=\varepsilon/3$ for an ultrarelativistic gas). The electric current is decomposed as
\be
I^\mu = \tilde{\rho}_\mathrm{e} u^\mu + j^\mu,
\ee
where $\tilde{\rho}_\mathrm{e} = - I^\mu u_\mu$ is the proper electric charge density and $j^\mu$ the \emph{conduction} current, normal to $u^\mu$ by construction. 
The electromagnetic fields are defined through the Faraday tensor and its dual as
\begin{align}
F^{\mu\nu} & = u^\mu e^\nu - u^\nu e^\mu + \epsilon^{\mu\nu\lambda\kappa} b_\lambda u_\kappa, \nonumber \\
F^{\star\mu\nu} & = u^\mu b^\nu - u^\nu b^\mu - \epsilon^{\mu\nu\lambda\kappa} e_\lambda u_\kappa,
\label{eq:f1}
\end{align}
where quantities are measured \emph{in the comoving frame of the fluid}. Thus, $e^\mu = F^{\mu\nu}u_\nu$ and $b^\mu = F^{\star\mu\nu}u_\nu$ are the electric and magnetic fields in the fluid rest frame, so that $e^\mu u_\mu = b^\mu u_\mu =0$ as well as $j^\mu u_\mu =0$. The electromagnetic contribution to the energy-momentum tensor can be expressed by using the $e^\mu$ and $b^\mu$ fields as
\be
T_\mathrm{f}^{\mu\nu} = (e^2 + b^2)u^\mu u^\nu + \tfrac{1}{2}(e^2+b^2)g^{\mu\nu} - e^\mu e^\nu - b^\mu b^\nu.
\label{eq:tf}
\ee
The rate of energy dissipation via Joule heating can be found by projecting Eq.~(\ref{eq:mf}) along the flow, then 
\be
- u_\nu \nabla_\mu T_\mathrm{m}^{\mu\nu} = \dot{\varepsilon} + (\varepsilon + p) \nabla_\mu u^\mu = j_\mu e^\mu,
\ee
vanishing in the ideal MHD as expected, where the dot indicates the $u^\mu\nabla_\mu$ time-like derivation along $u^\mu$.

For resistive plasmas, the relativistic Ohm's law is generally written as
\be
j^\mu = \sigma^{\mu\nu} e_\nu,
\label{eq:ohm3}
\ee
where $\sigma^{\mu\nu}$ is a tensor of electric conductivity, anisotropic in the most general case \citep{Bekenstein:1978}. In the isotropic case the relation is simply $j^\mu = \sigma e^\mu$, where the conductivity coefficient is the inverse of the resistivity $\eta$ introduced in the previous section. Notice that an evolutionary equation with a finite relaxation time should be actually introduced in order to avoid non-causal effects, as for the dissipative effects in extended irreversible hydrodynamics \citep{Pavon:1980}. In the ideal MHD limit $\sigma^{\mu\nu}\to\infty$, and in order to avoid divergent currents in the plasma, we simply assume that the comoving electric field vanishes, that is $e^\mu = 0$. Ohmic dissipation from Eq.~(\ref{eq:ohm3}) is $\sigma^{\mu\nu}e_\mu e_\nu$ in the general case, is proportional to $e^2$, or to $j^2$, in the isotropic case, and of course it is zero in the ideal case.

We turn now our attention to the chiral effect, exploiting the natural parallelism with the dynamo action. Following the same approach as in \citep{Bucciantini:2013}, it is natural to extend Eq.~(\ref{eq:ohm2}), valid for classical MHD, to the relativistic case by expressing the most general Ohm law valid for chiral-dynamo resistive MHD as
\be
j^\mu = \sigma_E e^\mu + \sigma_B b^\mu,
\label{eq:ohm4}
\ee
where the relations among the conduction current, the electric field and the magnetic field are assumed to hold in the comoving frame of the fluid. Here for simplicity we have assumed an isotropic tensor for the Ohmic conductivity (including eventually a mean-field turbulent contribution), that is $\sigma_{\mu\nu} = \sigma_E g_{\mu\nu}$, and again we have supposed that relaxation times are small compared to advection times. The second term is the one responsible for the CME, which in covariant form is naturally expressed as in Eq.~(\ref{eq:ohm4}), naturally reducing to the classical expression for low velocities.

\subsection{$3+1$ formulation}

In view of implementation of the system of our DC-GRMHD equations in numerical codes, the next necessary step is to move to the reference frame of the so-called \emph{Eulerian observer} of velocity $n^\mu$, rather than $u^\mu$, as needed to single out the time evolution. This leads to the so-called $3+1$ formulation \citep{Alcubierre:2008,Gourgoulhon:2012,Rezzolla:2013}. Any vector parallel to $n^\mu$ will be a time-like vector, while a vector which is normal to $n^\mu$ will be a spatial vector, to be treated with the standard three-metric (for which we will be using latin indices $i,j,\ldots =1,2,3$). In a Minkowskian spacetime we simply have $n^\mu=(1,0)$, thus basically indicating the standard laboratory frame. 

In the presence of gravity, and hence on a curved manifold, the $3+1$ form of the spacetime metric is usually expressed in terms of a scalar \emph{lapse function} $\alpha$, a spatial vector \emph{shift vector} $\beta^i$, and the three-metric $\gamma_{ij}$, that is
\be
\mathrm{d}s^2 = \! -\alpha^2\mathrm{d}t^2+
\gamma_{ij}\,(\mathrm{d}x^i\!+\beta^i\mathrm{d}t)(\mathrm{d}x^j\!+\beta^j\mathrm{d}t),
\ee
so that the Eulerian observer has unit vector
\be
n_\mu = (-\alpha, 0), \quad n^\mu = (1/\alpha, -\beta^i/\alpha),
\ee
reducing to the flat spacetime case when $\alpha=1$ and $\beta^i=0$.
The covariant derivative of $n^\mu$ can be split as
\be
\nabla_\mu n_\nu = - n_\mu \partial_\nu\log\alpha - K_{\mu\nu},
\ee
where $K_{\mu\nu}$ is the \emph{extrinsic curvature tensor} (symmetric and spatial), which is provided by the solution of Einstein equations together with $\gamma_{ij}$, given the gauge fields $\alpha$ and $\beta^i$.

In the $3+1$ formulation, the four velocity of the fluid is conveniently split as
\be
u^\mu = \Gamma n^\mu + \Gamma \varv^\mu,
\ee
where $\Gamma=-n_\mu u^\mu = 1/\sqrt{1-\varv^2}$ is the usual Lorentz factor for the three-velocity $v^i$, derived from the normalizing conditions $u^\mu u_\mu = n^\mu n_\mu = -1$ and $\varv^\mu n_\mu = 0$. 
The electromagnetic fields are split according to $n^\mu$ as usual like
\begin{align}
F^{\mu\nu} & = n^\mu E^\nu - n^\nu E^\mu + \epsilon^{\mu\nu\lambda\kappa} B_\lambda n_\kappa, \nonumber \\
F^{\star\mu\nu} & = n^\mu B^\nu - n^\nu B^\mu - \epsilon^{\mu\nu\lambda\kappa} E_\lambda n_\kappa,
\label{eq:f2}
\end{align}
where $E^\mu$ and $B^\mu$ are the standard spatial electric and magnetic fields, and similarly to Eq.~(\ref{eq:tf}) we may write
\be
T_\mathrm{f}^{\mu\nu} = (E^2 \! + \! B^2)n^\mu n^\nu + \tfrac{1}{2}(E^2 \! + \! B^2)\gamma^{\mu\nu} 
\! - \! E^\mu E^\nu \! - \! B^\mu B^\nu.
\label{eq:tf2}
\ee
The conserved total energy-momentum tensor can be equivalently split as
\begin{equation}
T^{\mu\nu} = \mathcal{E} n^\mu n^\nu + S^\mu n^\nu + S^\nu n^\mu + S^{\mu\nu}, 
\end{equation}
where $\mathcal{E} = T^{\mu\nu}n_\mu n_\nu$ is the total energy density as measured by the Eulerian observer, $S^i = - \gamma^i_{\,\mu}T^{\mu\nu}n_\nu$ is the momentum flux, and $S^{ij} = \gamma^i_{\,\mu}\gamma^j_{\,\nu}T^{\mu\nu}$ 
is the stress tensor. Using the expressions for $T_\mathrm{m}$ and $T_\mathrm{f}$ and that for $u^\mu$ we find
\begin{align}
\mathcal{E} & = (\varepsilon + p) \Gamma^2 - p + u_\mathrm{em}, \nonumber \\
S_i & = (\varepsilon + p) \Gamma^2 v_i + \epsilon_{ijk} E^j B^k, \\
S_{ij} & = (\varepsilon + p) \Gamma^2 v_i v_j -E_i E_j - B_i B_j + (p + u_\mathrm{em}) \gamma_{ij}, \nonumber
\end{align}
where $u_\mathrm{em}=\tfrac{1}{2}(E^2+B^2)$ and $\epsilon_{ijk} = \sqrt{\gamma}[ijk]$ is the Levi-Civita pseudo-tensor for the three-metric, $\gamma = \mathrm{det}(\gamma_{ij})$, and $[ijk]$ is the non-dimensional Levi-Civita symbol.

Consider now the four-current. In the $3+1$ split this can be expressed as
\be
I^\mu = \rho_\mathrm{e} n^\mu + J^\mu,
\ee
where $\rho_\mathrm{e}=-n_\mu I^\mu$ is the electric charge density and $J^\mu$ the spatial current, as measured by the Eulerian observer.  By equating the two relations for $I^\mu$ and using the definition of $u^\mu$, the conduction current becomes
\be
j^\mu = (\rho_\mathrm{e}-\tilde{\rho}_\mathrm{e}\Gamma) n^\mu + J^\mu - \tilde{\rho}_\mathrm{e}\Gamma \varv^\mu,
\label{eq:j}
\ee
to be plugged in Ohm's law. The missing ingredients are the electromagnetic fields in the comoving frame $e^\mu$ and $b^\mu$, to be expressed in $3+1$ form as well. By using Eqs.~(\ref{eq:f1}) and (\ref{eq:f2}) we find 
\begin{align}
e^\mu  = \, F^{\mu\nu} u_\nu = &
\Gamma [(\varv^\nu E_\nu)n^\mu + E^\mu + \epsilon^{\mu\nu\lambda}\varv_\nu B_\lambda] , \nonumber \\
b^\mu  = \!  F^{*\mu\nu} u_\nu = &
\Gamma [(\varv^\nu B_\nu)n^\mu + B^\mu - \epsilon^{\mu\nu\lambda}\varv_\nu E_\lambda] ,
\end{align}
where $\epsilon^{\mu\nu\lambda\kappa}n_\kappa\equiv\epsilon^{\mu\nu\lambda}$. The above expressions, together with Eq.~(\ref{eq:j}), are then inserted into the covariant form of our generalized Ohm's law Eq.~(\ref{eq:ohm4}). By retrieving $\rho_\mathrm{e}\Gamma$ from the time component, the spatial part becomes
\begin{align}
J^i = \rho_\mathrm{e} v^i 
& + \sigma_E \Gamma [ E^i + \epsilon^{ijk}v_jB_k - (v_kE^k)v^i]  \nonumber \\
& + \sigma_B \Gamma [ B^i - \epsilon^{ijk}v_jE_k + (v_kB^k)v^i] .
\label{eq:ohm5}
\end{align}
The above expression (\ref{eq:ohm5}) for the spatial current is the final $3+1$ form of our novel Ohm's law for dynamo-chiral resistive relativistic MHD, and it applies unchanged to both flat or curved spacetimes, thus the above form is perfectly valid for the most general case of full GRMHD.

Notice that if the additional constraint of a vanishing comoving charge density were imposed, $\tilde{\rho}_\mathrm{e}=0 \Rightarrow I^\mu \equiv j^\mu$, all terms proportional to $v^i$ would vanish in Eq.~(\ref{eq:ohm5}), but then $\rho_\mathrm{e}$ would have two conflicting definitions, one as the divergence of $E^i$ from Maxwell's equations (see below) and one from the time component of Eq.~(\ref{eq:ohm4}).

The set of evolution equations for resistive GRMHD in conservative form is found by splitting Eqs.~(\ref{eq:hydro}) and (\ref{eq:maxwell}) \citep{Bucciantini:2013,Dionysopoulou:2013}, and we find the system of nonlinear equations
\begin{align}
 \partial_t (\sqrt{\gamma} D) & +  \partial_k [ \sqrt{\gamma} ( \alpha D v^k - \beta^k D)]  = 0, \nonumber \\
\partial_t (\sqrt{\gamma} S_i) & +  \partial_k [ \sqrt{\gamma} ( \alpha S^k_{\,i} - \beta^kS_i)]  = \nonumber \\
& \sqrt{\gamma}(\tfrac{1}{2}\alpha S^{lm}\partial_i\gamma_{lm} + 
S_k\partial_i\beta^k - \mathcal{E}\partial_i\alpha), \nonumber \\
\partial_t (\sqrt{\gamma} \mathcal{E}) & +  \partial_k [ \sqrt{\gamma} ( \alpha S^k - \beta^k \mathcal{E})]  =  \nonumber \\
& \sqrt{\gamma}(\alpha S^{lm}K_{lm} - S^k \partial_k\alpha), \\
\partial_t (\sqrt{\gamma} B^i) & + [ijk] \partial_j ( \alpha E_k + [klm] \sqrt{\gamma} \beta^l B^m) = 0, \nonumber \\
\partial_t (\sqrt{\gamma} E^i) & -  [ijk] \partial_j ( \alpha B_k - [klm] \sqrt{\gamma} \beta^l E^m) =  \nonumber \\
& -\sqrt{\gamma}(\alpha J^i - \beta^i \rho_\mathrm{e}), \nonumber
\end{align}
where $D=\rho\Gamma$ is the mass density measured by the Eulerian observer. The metric terms should be provided by the solution of Einstein equations. As an alternative, for a given metric (even time-dependent), the term with the extrinsic curvature can be also expressed as
\be
\alpha S^{lm}K_{lm} = \tfrac{1}{2}S^{lm}(\beta^k\partial_k\gamma_{lm} - \partial_t\gamma_{lm}) + S^l_{\,m}\partial_l\beta^m.
\ee
Notice that the first three hydrodynamics equations contain fluxes in the standard divergence form, while Maxwell equations in curl form. This fact is related to the presence of the two non-evolutionary constraints
\be
\partial_k ( \sqrt{\gamma} B^k) = 0, \qquad \partial_k ( \sqrt{\gamma} E^k) = \sqrt{\gamma} \, \rho_\mathrm{e},
\ee
and while the solenoidal constraint for $B^i$ is analytically (but not numerically, especially for shock-capturing schemes) preserved during evolution, the second is used to define the charge $\rho_\mathrm{e}$ in both Ohm's law and in the equation for $E^i$. The above GRMHD set is then a system of 11 evolution equations for the 11 \emph{conservative} variables $[ D, S_i, \mathcal{E}, B^i, E^i ]$, which is closed by an equation of state of the form $p=\mathcal{P}(\rho,\varepsilon)$ and by the generalized Ohm law Eq.~(\ref{eq:ohm5}).

In the simplest case of a Minkowskian flat spacetime, the set of dynamo-chiral resistive (special) relativistic MHD equations do not contain the metric terms, and they can be also expressed in the more familiar vector form as
\begin{align}
& \partial_t D + \bm{\nabla}\cdot ( \rho \Gamma \bm{v}) = 0, \nonumber \\
& \partial_t \bm{S} + \bm{\nabla}\cdot [(\varepsilon + p) \Gamma^2 \bm{v}\bm{v} + (p + u_\mathrm{em}) \bm{I} -  \bm{E}\bm{E} - \bm{B}\bm{B}  ] = 0, \nonumber \\
& \partial_t \mathcal{E} + \bm{\nabla}\cdot [ (\varepsilon + p) \Gamma^2 \bm{v} + \bm{E}\times\bm{B} ] = 0, \\
& \partial_t \bm{B}  + \bm{\nabla}\times\bm{E} = 0, \nonumber \\
& \partial_t \bm{E}  - \bm{\nabla}\times\bm{B} = - \bm{J}, \nonumber 
\end{align}
where the hydrodynamical conserved variables are $D=\rho\Gamma$, $\bm{S}=(\varepsilon + p) \Gamma^2 \bm{v} + \bm{E}\times\bm{B}$, and $\mathcal{E}=(\varepsilon + p) \Gamma^2 -  p + u_\mathrm{em}$. The remaining Maxwell equations are the constraints
\be
\bm{\nabla}\cdot\bm{B}=0, \qquad 
\bm{\nabla}\cdot\bm{E}=\rho_\mathrm{e},
\ee
the latter to be used in Ohm's law
\begin{align}
\bm{J} = \rho_\mathrm{e} \bm{v} 
& + \sigma_E \Gamma [\bm{E} + \bm{v}\times\bm{B} - (\bm{v}\cdot\bm{E})\bm{v}]  \nonumber \\
& + \sigma_B \Gamma [\bm{B} - \bm{v}\times\bm{E} - (\bm{v}\cdot\bm{B})\bm{v}],
\end{align} 
which in spite of the bold face aspect for vectors, is precisely equivalent to Eq.~(\ref{eq:ohm5}).

\section{Numerical test: magnetar model}
\label{sect:numerical} 

\begin{figure*}
\includegraphics[width=\columnwidth]{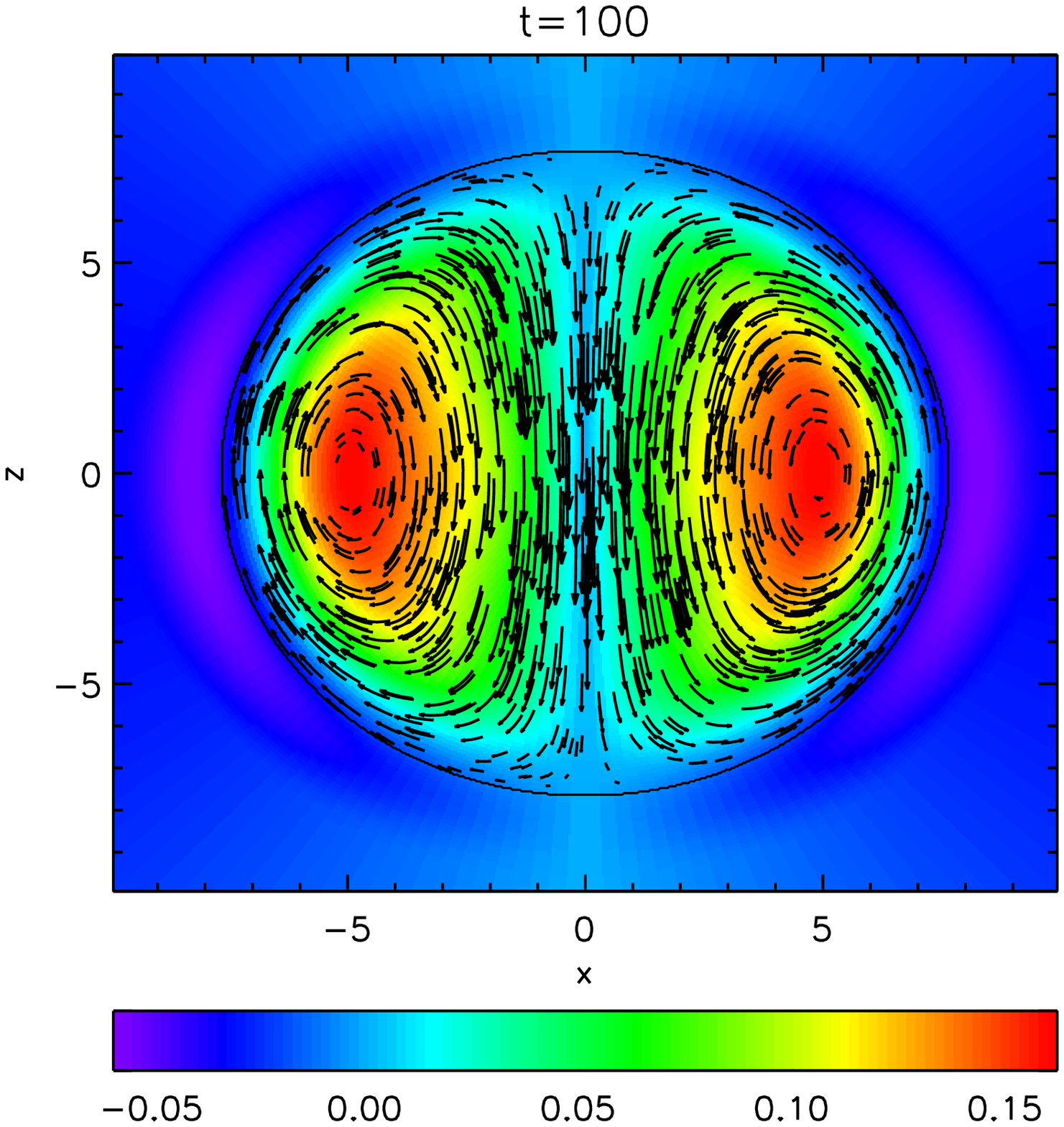}
\includegraphics[width=\columnwidth]{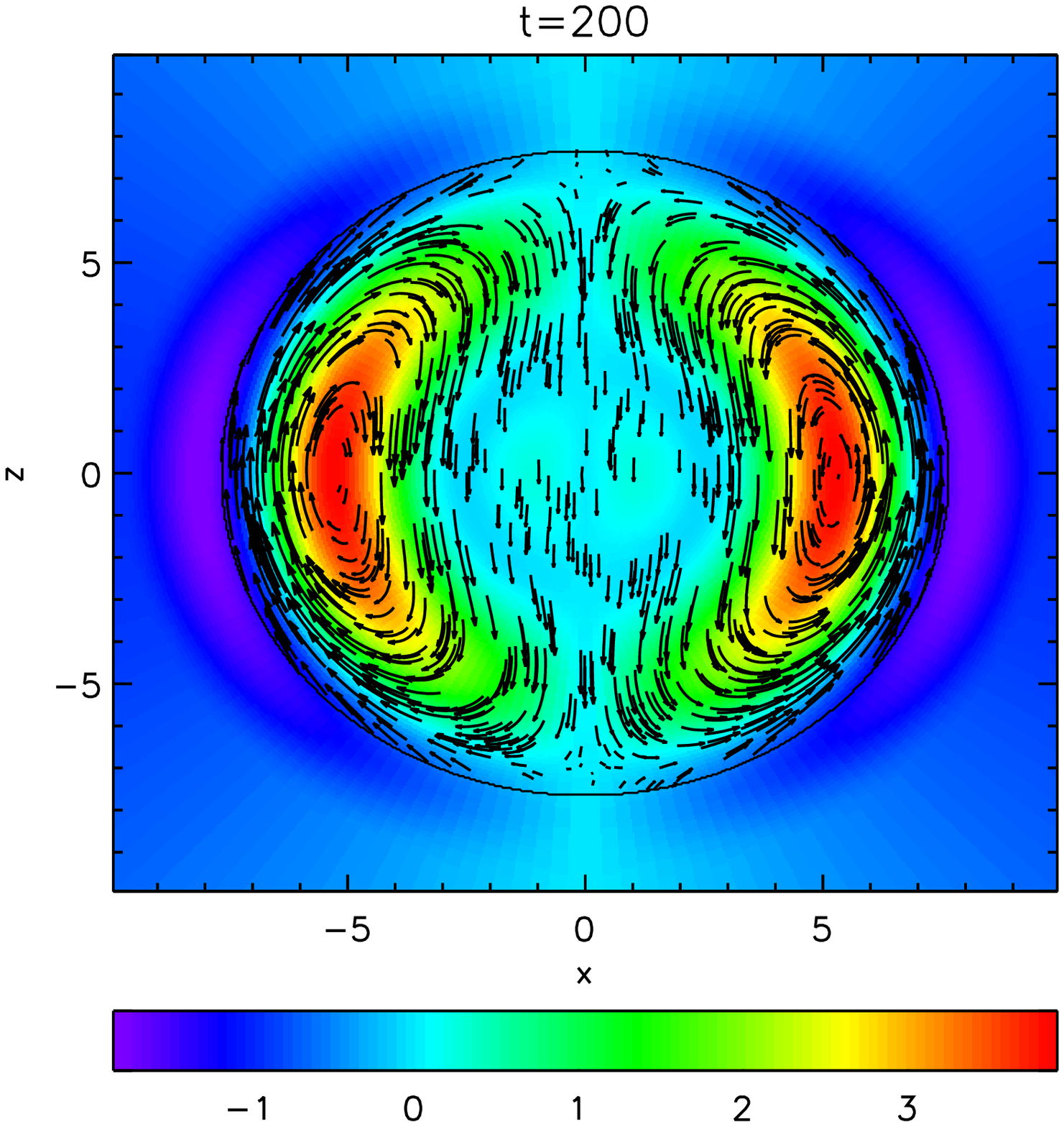}
\caption{DC-GRMHD magnetar model. The magnetic field configuration at two different times of evolution, $t=100$ and $t=200$. Fieldlines refer to the poloidal component, whereas color contours to the toroidal one, with values expressed in units of $10^{15}$~G.}
\label{fig:NS}
\end{figure*}

For numerical implementation into the \texttt{ECHO} (\emph{Eulerian Conservative High-Order}) code \citep{Del-Zanna:2007,Bucciantini:2011a} and extensive testing, the reader is referred to \cite{Bucciantini:2013,Del-Zanna:2014}. Here it is sufficient to notice that due to the presence of possibly large values of $\sigma_E$ and $\sigma_B$ (both contain a resistivity $\eta$ term in the denominator) in Ohm's law, \emph{stiff} source terms arise in the last equation for the evolution of $\bm{E}$, so that specific numerical methods (e.g. implicit time integration) must be employed. When the dynamo-chiral coefficient $\sigma_B$ can be neglected, the set of equations described in the previous section reduces to that for resistive relativistic MHD in Minkowski metric, see \cite{Komissarov:2007,Palenzuela:2009,Del-Zanna:2016a} for numerical methods and applications.

In this section we present an example of solution of the DC-GRMHD equations in a case where strong gravity is important, namely the growth of an initial magnetic field inside a neutron star as a simplified model for a (proto)-magnetar. This test was already presented in \cite{Bucciantini:2013} where the driving force was the classical dynamo, here we re-run it in the light of the (laminar) CME. The spacetime metric is calculated assuming conformal flatness, theoretical details and the methods employed for the numerical implementation in the \texttt{ECHO} code can be found in  \citep{Bucciantini:2011a}. The initial neutron star structure is built using the freely available \texttt{XNS} tool, which can be applied to very different configurations of density, velocity and magnetic field \citep{Pili:2014,Pili:2015,Pili:2017,Bucciantini:2015,Del-Zanna:2018}.

The initial configuration is a non-rotating star with central density $\rho_c  = 1.28\times 10^3$ (in geometrized units $c=G=M_\odot = 1$) and polytropic equation of state $p=K \rho^{1+1/n}$ (assuming $K=100$ and $n=1$, with an initial purely toroidal field (here in code units)
\be
B_\mathrm{T} = \sqrt{B_\phi B^\phi} = 10^{-4} \alpha\psi^2\, r\sin\theta\, \rho h,
\ee
with $h = 1 + (1+n) K \rho^{1/n} $ the relativistic specific enthalpy.
Here $\alpha$ is the lapse function and $\psi$ the conformal factor, entering the $3+1$ conformally flat metric
\be
ds^2 = - \alpha^2 dt^2 + \psi^4 (dr^2 +r^2d\theta^2 + r^2\sin^2\!\theta\, d\phi^2).
\ee
The electrical and chiral conductivities are considered constant and uniform in the whole star and are set to $\sigma_E = 20$ and $\sigma_B = -2$ respectively, in order to match the same values employed by \cite{Bucciantini:2013} for the mean-field dynamo test. Here we do not consider the possible time evolution of the chiral chemical potential and conductivity, for simplicity. The simulation is performed in spherical-like coordinates in a domain $r=[0,10]$, $\theta=[-\pi,\pi]$ assuming axial symmetry, that is invariance in $\phi$, up to $t=200$. We employ 100 points in the radial direction and 80 along $\theta$, this low resolution is enough to capture the instability growth and the test can be run on a simple laptop. The DC-GRMHD equations are solved together with Einstein equations, though the initial equilibrium is only slightly affected by the growing magnetic field, so results are very similar to those obtained in the Cowling approximation of a fixed spacetime metric.

\begin{figure}
\includegraphics[width=\columnwidth]{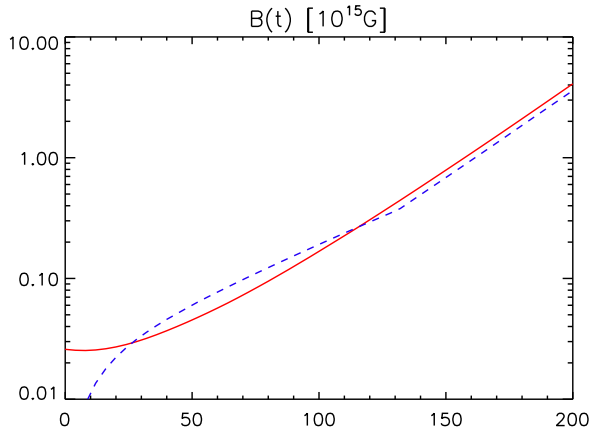}
\caption{DC-GRMHD magnetar model. The exponential growth in time of the magnetic field toroidal component $B_T$ (red solid line) and poloidal component $B_P$ (blue dashed line) due to the CME.}
\label{fig:B}
\end{figure}

In Fig.~\ref{fig:NS} we show the neutron star at half ($t=100$) and maximum ($t=200$) times of evolution, with fieldlines of the poloidal component and strength of the toroidal one in units of $10^{15}$~G, as indicated by the colors. Notice that the magnetic field changes its shape during evolution, and above all the CME manages to amplify the toroidal component from the initial low value, up to $B_\mathrm{T} \sim 10^{13}$~G at $t=100$ and to $B_\mathrm{T} \simeq 4 \times 10^{15}$~G at $t=200$. Such rapid evolution can be better appreciated in Fig.~\ref{fig:B}, where the growth of both the toroidal and poloidal components is shown in logarithmic scale as a function of time. Notice the nearly exponential behavior for both components, typical of the CME, and the fact that a single dominant mode is present for the toroidal component, whereas the poloidal component $B_\mathrm{P}$ shows an initial growth followed by an exponential behaviour for a different eigenmode, with a slower rate, then relaxing around $t\simeq 140$ to the same one responsible for the growth of $B_\mathrm{T}$. For this we estimate a growth rate $\gamma\simeq 0.034$. The value is more or less what expected for a laminar dynamo for which $\gamma\sim \alpha_\mathrm{dyn} / \lambda$, using $\alpha_\mathrm{dyn}=| \sigma_B |/\sigma_E =0.1$ and estimating $\lambda \sim |\bm{B}| /|\nabla\bm{B} |\simeq 3$ from the figures for the most unstable eigenmode. Notice that times are in geometrized code units and the value of $\sigma_B\propto\gamma$ has been selected in order to show a sufficient growth within the chosen evolution time (realistic values of $\sigma_B$ would certainly require a much longer evolution).

Further possible applications of chiral/dynamo instabilities in the GRMHD regime where strong gravity is needed, other than the interior of magnetars, may be accretion disks orbiting in the close vicinities of black holes. The growth of the magnetic field in the kinematical regime for various configurations of the equilibrium disk, of the initial seed field, and of the dynamo and dissipative coefficients were presented by \cite{Bugli:2014}, whereas the interplay of the other typical disk instabilities (Papaloizou-Pringle and magneto-rotational) in full 3D simulations have been recently investigated in detail by \cite{Bugli:2018}.

\section{Conclusions}

The Chiral Magnetic Effect (CME), that is the quantum phenomenon of creation of an equilibrium current in a plasma of imbalanced chiral fermions, is believed to be important in many astrophysical scenarios where the generation and amplification of magnetic fields is crucial. These may range from the primordial fireball in the early Universe to the quark-gluon plasma possibly present in the interior of proto-neutron stars. 

In the present paper we have discussed the analogy between the CME and the dynamo action for relativistic plasmas in the GRMHD approximation \citep{Bucciantini:2013}, and proposed a unified formalism for the covariant equations based on the splitting according to the bulk plasma fluid four-velocity $u^\mu$. In particular, the generalized Ohm's law involving the conduction current and the electric and magnetic fields must be written in the comoving frame, and we have done so by using conductivity coefficients: the term $j^\mu \propto b^\mu$ being the chiral-dynamo one and $j^\mu \propto e^\mu$ the resistive one. For numerical relativity applications, the covariant equations have been translated into the $3+1$ formalism, deriving for the first time the DC-GRMHD system valid both in flat and curved spacetime. In the latter case geometrical terms arise in the conservative variables, fluxes and source terms, to be found from the Einstein field equations (or prescribed).  

An aspect worth of investigation is the relationship between the CME and magnetic reconnection processes. It is known that the axial chemical potential $\mu_\mathrm{A}$ responsible for the $\bm{J}_\mathrm{CME}\propto\bm{B}$ conduction current obeys an evolution equation sourced by an anomaly term $\propto \bm{E}\cdot\bm{B}$. This is the same non-ideal term arising in resistive layers where reconnection processes are important, and it has been recently argued that the CME may be directly induced by reconnection processes, even in absence of an initial chiral imbalance \citep{Hirono:2016}. In the case of fully relativistic plasmas, the reconnection process arising from the \emph{tearing instability} of a thin current sheet has been recently investigated by means of analytical and numerical modeling, for the first time within relativistic MHD, in \cite{Del-Zanna:2016a}. It was shown that fast (efficient) reconnection can be easily achieved, as it is required to explain various high-energy astrophysical scenarios where explosive events are induced by a sudden release of magnetic energy, like for magnetar eruptions \citep{Lyutikov:2006a} or gamma-ray flares in the Crab nebula \citep{Cerutti:2014a,Del-Zanna:2016}. The complex interplay between these phenomena and the CME induced by reconnection itself will be subject of future work.

However, we believe that the primary field of application of the present study will be the investigation of the origin of the high magnetic fields of (proto)-magnetars, either produced in core collapse or binary merger events. Newly born and fast spinning proto-magnetars have gained increasingly more attention since they are nowadays considered to be the best candidates for the engine of both long and short GRBs, including the kilonova ejecta produced in the merger \citep{Metzger:2011,Bucciantini:2012,Rowlinson:2013,Metzger:2018}. Such compact objects are characterized by strong gravity and matter above nuclear density, so that a fully relativistic treatment is required, and in section \ref{sect:numerical} we have shown that the \texttt{ECHO} code is already capable of solving the DC-GRMHD equations, even with dynamical spacetime, in such an environment.

Future work will be devoted to the investigation of the CME in realistic magnetar models, including a more appropriate equation of state which could also be that of a \emph{quark star} with a core made up by QGP, as required by recent GRB models \citep{Drago:2016,Pili:2016}. Depending on the temperature, in the dense cores of proto-neutron stars conditions favourable to the CME could be met \citep{Sigl:2016}. If one assumes $\mu_A\propto \mu_e$ (left-chiral electrons have all turned into neutrinos and the chirality \emph{flipping} rate is negligible), an estimate for the field that could be produced within a few seconds is \citep{Schober:2018}
\be
B_\mathrm{max} \simeq 1.2\times 10^{12}\,\mathrm{G} 
\left(\frac{\mu_e}{250~\mathrm{MeV}}\right)^{3/2}\left(\frac{\lambda}{1~\mathrm{cm}}\right)^{-1/2},
\ee
where a standard value of the Fermi energy for electrons $\mu_e \gg k_\mathrm{b}T$ has been used and $\lambda$ is the typical correlation scale of the turbulence induced by the initial small-scale chiral dynamo. Larger values of $\mu_A$ may be actually needed to reach the $B$ inside magnetars, though for $\mu_A\sim k_\mathrm{b}T$ the CME instability scale may become smaller than the mean-free path of the plasma, and the MHD description should be replaced by a kinetic one.

\section*{Acknowledgements}

The authors are grateful to D.~Kharzeev, F.~Becattini and G.~Inghirami for stimulating discussions, and to the anonymous referee for helpful suggestions. The authors acknowledge support from the PRIN-MIUR project \emph{Multi-scale Simulations of High-Energy Astrophysical Plasmas} (Prot.~2015L5EE2Y) and from the INFN - TEONGRAV initiative (local PI: LDZ). NB has been supported by a EU FP7 - CIG grant issued to the NSMAG project (PI: NB).

\bibliographystyle{mn2e}

\end{document}